\documentclass[journal, 10pt, journal, twoside]{IEEEtran}
\usepackage{cite}

\ifCLASSINFOpdf
   \usepackage[pdftex]{graphicx}
\else
   \usepackage[dvips]{graphicx}
\fi
\usepackage{amsmath}
\usepackage{url}

\usepackage{amsfonts,amssymb}
\usepackage{bm}
\usepackage{listings}
\usepackage{float}
\usepackage{comment}
\usepackage{hyperref}

\newtheorem{theorem}{Theorem}

\newtheorem{lemma}{Lemma}
\newtheorem{prop}{Proposition}
\newtheorem{definition}{Definition}
\newtheorem{notations}{Notations}

\newtheorem{remark}{Remark}

\newcommand{\argmax}{\mathop{\rm arg~max}\limits}

\begin{document}
%
\title{
Reinforcement Learning of Control Policy for Linear Temporal Logic Specifications Using Limit-Deterministic Generalized B\"{u}chi Automata
}
%
%
%

\author{Ryohei~Oura, 
        Ami~Sakakibara,~\IEEEmembership{Student~Member,~IEEE,}
        and~Toshimitsu~Ushio,~\IEEEmembership{Member,~IEEE}
\thanks{
This work was partially supported by JST-ERATO HASUO Project Grant Number JPMJER1603, Japan, and JST-Mirai Program Grant Number JPMJMI18B4, Japan.
The work of A.~Sakakibara was supported by the Grant-in-Aid for Japan Society for the Promotion of Science Research Fellow under Grant JP19J13487.
}
\thanks{
The authors are with the Graduate School of Engineering Science, Osaka University, Toyonaka 560-8531, Japan (e-mail: {r-oura, sakakibara}@hopf.sys.es.osaka-u.ac.jp; ushio@sys.es.osaka-u.ac.jp).
}
}

%
%

\markboth{Journal of \LaTeX\ Class Files,~Vol.~14, No.~8, August~2015}%
{Shell \MakeLowercase{\textit{et al.}}: Bare Demo of IEEEtran.cls for IEEE Journals}
%



\maketitle
\thispagestyle{empty}
\begin{abstract}
This letter proposes a novel reinforcement learning method for the synthesis of a control policy satisfying a control specification described by a linear temporal logic formula.  We assume that the controlled system is modeled by a Markov decision process (MDP).
We convert the specification to a limit-deterministic generalized B\"{u}chi automaton (LDGBA) with several accepting sets that accepts all infinite sequences satisfying the formula.
The LDGBA is augmented so that it explicitly records the previous visits to accepting sets.
We take a product of the augmented LDGBA and the MDP, based on which we define a reward function. The agent gets rewards whenever state transitions are in an accepting set that has not been visited for a certain number of steps.
Consequently, sparsity of rewards is relaxed and optimal circulations among the accepting sets are learned. We show that the proposed method can learn an optimal policy when the discount factor is sufficiently close to one.
\end{abstract}

\begin{IEEEkeywords}
Reinforcement Learning, Linear Temporal Logic, Limit-Deterministic B\"{u}chi Automata.
\end{IEEEkeywords}

%
\IEEEpeerreviewmaketitle

\section{Introduction}
%
%
%
%

\IEEEPARstart{R}{einforcement} learning (RL) \cite{Sutton} is a useful approach to learning an optimal policy from sample behaviors of a controlled system with inherent stochasticity, e.g., a Markov decision process (MDP) \cite{Puterman}, when the probabilities associated with the controlled system are unknown a priori. 
In RL, we use a reward function that assigns a reward to each transition in the behaviors and evaluate a control policy by the return, namely an expected (discounted) sum of the rewards along the behaviors.
One of the recent trends is to apply RL to synthesis problems under linear temporal logic (LTL) constraints.
LTL is a formal language with rich expressivity and thus suitable for describing complex missions assigned to a controlled system \cite{BK2008,Belta2017}. 

It is known that any LTL formula can be converted into an $\omega$-automaton with a B\"{u}chi or a Rabin acceptance condition \cite{BK2008}. 
In many studies on LTL synthesis problems using RL, 
reward functions are formed systematically from automata corresponding to the LTL specification. 
This direction was first investigated by Sadigh et al. \cite{Sadigh2014}, where they defined a reward function based on the acceptance condition of a deterministic Rabin automaton \cite{BK2008} that accepts all words satisfying the LTL constraint. 
Reward functions defined on specification automata were also proposed for a deep reinforcement learning method \cite{QDYYM2019} and for an extension of LTL in collaboration with a control barrier function \cite{Li2019}.

Recently, a limit-deterministic B\"{u}chi automaton (LDBA) or a generalized one (LDGBA) is paid much attention to as an $\omega$-automaton corresponding to the LTL specification \cite{SEJK2016}.
The RL-based approaches to the synthesis of a control policy using LDBAs or LDGBAs have been proposed in \cite{HAK2019,HKAKPL2019,Hahn2019,BWZP2019}.
An (LD)GBA has multiple accepting sets and accepts behaviors visiting all accepting sets infinitely often. 
One possible approach to generalized B\"{u}chi acceptance conditions 
is to convert a GBA into a corresponding BA, which has a single accepting set.
The conversion, however, fixes the order of visits to accepting sets of the GBA \cite{BK2008} and causes the sparsity of the reward, which is a critical issue in RL-based controller synthesis.
Another approach to RL-based synthesis for generalized B\"{u}chi conditions is the accepting frontier function introduced in \cite{HAK2019,HKAKPL2019}, based on which the reward function is defined. 
However, the function is memoryless, that is, it does not provide information of accepting sets that have been visited, which is important to improve learning performance.

In this letter, we propose a novel method to augment an LDGBA converted from a given LTL formula.
Then, we define a reward function based on the acceptance condition of the product MDP of the augmented LDGBA and the controlled system represented as the MDP.
The rest of the letter is organized as follows.
Section II reviews an MDP, LTL, and automata.
Section III proposes a novel RL-based method for the synthesis of a control policy.
Section IV presents a numerical example to show the effectiveness of our proposed method.


\begin{notations}
  $\mathbb{N}_0$ is the set of nonnegative integers. $\mathbb{R}_{\geq 0}$ is the set of nonnegative real numbers.
  For sets $A$ and $B$, $AB$ denotes their concatenation, i.e., $AB = \{ ab ; a \in A, b \in B \}$. $A^{\omega}$ denotes the infinite concatenation of the set $A$ and $A^{\ast}$ denotes the finite one, i.e., $A^{\omega} = \{ a_0a_1\ldots ; a_n \in A, n \in \mathbb{N}_0 \} $ and $A^{\ast} = \{ a_0a_1\ldots a_n ; a_n \in A, n \in \mathbb{N}_0 \} $, respectively.
  $\varepsilon \in A^*$ is the empty string. 
\end{notations}

\section{Preliminaries}

\subsection{Markov Decision Process}

\begin{definition}
A (labeled) Markov decision process (MDP) is a tuple $M$ = $(S, A, P, s_{init}, AP, L)$, where S is a finite set of states, $A$ is a finite set of actions, $P:S \times S \times A \rightarrow [0,1]$ is a transition probability function, $s_{init} \in S$ is the initial state, $AP$ is a finite set of atomic propositions, and $L : S \times A \times S\ \rightarrow\ 2^{AP}$ is a labeling function that assigns a set of atomic propositions to each transition. Let $\mathcal{A}(s) = \{ a \in A ; \exists s^{\prime} \in S \text{ s.t. } P(s^{\prime} | s,a) \neq 0 \}$. Note that $\sum_{s' \in S} P(s'|s,a) = 1$ for any state $s \in S$ and action $a \in \mathcal{A}(s)$.

In the MDP $M$, an infinite path starting from a state $s_0 \in S$ is defined as a sequence $\rho\ =\ s_0a_0s_1 \ldots\ \in S (A S)^{\omega}$ such that $P(s_{i+1}|s_i, a_i) > 0$ for any $ i \in \mathbb{N}_0$. A finite path is a finite sequence in $S (A S)^{\ast}$. In addition, we sometimes represent $\rho$ as $\rho_{init}$ to emphasize that $\rho$ starts from $s_0 = s_{init}$.
For a path $\rho\ =\ s_0a_0s_1 \ldots$, we define the corresponding labeled path $L(\rho)\ =\ L(s_0,a_0,s_1)L(s_1,a_1,s_2) \ldots \in (2^{AP})^{\omega}$. 
$InfPath^{M}\ ( \text{resp., }FinPath^{M})$ is defined as the set of infinite (resp., finite) paths starting from $s_0=s_{init}$ in the MDP $M$. 
For each finite path $\rho$, $last(\rho)$ denotes its last state.
\end{definition}

\begin{definition}
  A policy on an MDP $M$ is defined as a mapping $\pi:FinPath^{M} \times \mathcal{A}(last(\rho)) \rightarrow [0,1]$. A policy $\pi$ is a {\it positional} policy if for any $ \rho \in FinPath^{M}$ and any $ a \in \mathcal{A}(last(\rho))$, it holds that $\pi(\rho, a)=\pi(last(\rho),a)$; and 
  there exists one action $ a' \in \mathcal{A}(last(\rho))$ such that $\pi(\rho, a) = 1$ if $a=a^{\prime}$, and $\pi(\rho, a) = 0$ for any $ a \in \mathcal{A}(last(\rho)) $ with $a\neq a'$.
\end{definition}

Let $InfPath^{M}_{\pi}$ (resp., $FinPath^{M}_{\pi}$) be the set of infinite (resp., finite) paths starting from $s_0=s_{init}$ in the MDP $M$ under a policy $\pi$. The behavior of the MDP $M$ under a policy $\pi$ is defined on a probability space $(InfPath^{M}_{\pi}, \mathcal{F}_{InfPath^{M}_{\pi}}, Pr^{M}_{\pi})$. 

A Markov chain induced by the MDP $M$ with a positional policy $\pi$ is a tuple $MC_{\pi} = (S_{\pi},P_{\pi},s_{init},AP,L)$, where $S_{\pi} = S$, $P_{\pi}(s'|s) = P(s'|s,a)$ for $s, s^{\prime} \in S$ and $a \in \mathcal{A}(s)$ such that $\pi(s,a) = 1$.
The state set $S_{\pi}$ of $MC_{\pi}$ can be represented as a disjoint union of a set of transient states $T_{\pi}$ and closed irreducible sets of recurrent states $R^j_{\pi}$ with $j \in \{ 1, \ldots ,h \}$, i.e., $ S_{\pi} = T_{\pi} \cup R^1_{\pi} \cup \ldots \cup R^h_{\pi} $ \cite{ESS}.
In the following, we say a ``recurrent class'' instead of a ``closed irreducible set of recurrent states'' for simplicity.

In the MDP $M$, we define a reward function $R:S \times A \times S \rightarrow \mathbb{R}_{\geq 0}$. The function returns the immediate reward received after the agent performs an action $a$ at a state $s$ and reaches a next state $s'$ as a result.

\begin{definition}
  For a policy $\pi$ on an MDP $M$, any state $s \in S$, and a reward function $R$, we define the expected discounted reward as
  \begin{align*}
    V^{\pi}(s)= \mathbb{E}^{\pi}[\sum_{n=0}^{\infty}\gamma^n R(S_n, A_n, S_{n+1})|S_0 = s],
  \end{align*}
where $\mathbb{E}^{\pi}$ denotes the expected value given that the agent follows the policy $\pi$ from the state $s$ and $\gamma \in [0,1)$ is a discount factor. The function $V^{\pi}(s)$ is often referred to as a state-value function under the policy $\pi$. For any state-action pair $(s,a) \in S \times A$, we define an action-value function $Q^{\pi}(s,a)$ under the policy $\pi$ as follows.
  \begin{align*}
    Q^{\pi}(s,a)= \mathbb{E}^{\pi}[\sum_{n=0}^{\infty}\gamma^n R(S_n, A_n, S_{n+1})|&S_0 = s, A_0 = a].
  \end{align*}
\end{definition}

\begin{definition}
  For any state $s \in S$, a policy $\pi^{\ast}$ is optimal if
  \begin{align*}
    \pi^{\ast} \in \argmax_{\pi \in \Pi^{pos}} V^{\pi}(s),
  \end{align*}
where $\Pi^{pos}$ is the set of positional policies over the state set $S$.
\end{definition}

\subsection{Linear Temporal Logic and Automata}

In our proposed method, we use linear temporal logic (LTL) formulas to describe various constraints or properties and to systematically assign corresponding rewards.
LTL formulas are constructed from a set of atomic propositions, Boolean operators, and temporal operators. We use the standard notations for the Boolean operators: $\top$ (true), $\neg$ (negation), and $\land$ (conjunction).
LTL formulas over a set of atomic propositions $AP$ are defined as
\begin{align*}
  \varphi ::=\top\ |\ \alpha \in AP\ |\ \varphi_1 \land \varphi_2\ |\ \neg \varphi\ |\ \text{{\bf X}} \varphi\ |\ \varphi_1 \text{{\bf U}} \varphi_2,
\end{align*}
where $\varphi$, $\varphi_1$, and $\varphi_2$ are LTL formulas.
Additional Boolean operators are defined as $\perp := \neg \top $, $\varphi_1 \lor \varphi_2 := \neg(\neg \varphi_1 \land \neg \varphi)$, and $\varphi_1 \Rightarrow \varphi_2 := \neg \varphi_1 \lor \varphi_2$.
The operators {\bf X} and {\bf U} are called ``next" and ``until", respectively.
Using the operator {\bf U}, we define two temporal operators: (1) {\it eventually}, $\text{{\bf F}} \varphi := \top \text{{\bf U}} \varphi $ and (2) {\it always}, $\text{{\bf G}} \varphi := \neg \text{{\bf F}} \neg \varphi$.

\begin{definition}
	For an LTL formula $\varphi$ 
	and an infinite path $\rho = s_0a_0s_1 \ldots$ of an MDP $ M $ with $ s_0 \in S $, the satisfaction relation $M,\rho \models \varphi$ is recursively defined as follows. 
	\begin{alignat}{2}
	& M, \rho \models \top,\nonumber \\
	& M, \rho \models \alpha \in AP &&\Leftrightarrow \alpha \in L(s_0,a_0,s_1),\nonumber \\
	& M, \rho \models \varphi_1 \land \varphi_2 &&\Leftrightarrow M, \rho \models \varphi_1 \land M, \rho \models \varphi_2,\nonumber \\
	& M, \rho \models \neg \varphi &&\Leftrightarrow M, \rho \not\models \varphi,\nonumber \\
	& M, \rho \models \text{{\bf X}}\varphi &&\Leftrightarrow M, \rho[1:] \models \varphi,\nonumber \\
	& M, \rho \models \varphi_1 \text{{\bf U}} \varphi_2 &&\Leftrightarrow \nonumber \\
	& \quad \exists j \geq 0, \ M, \rho[j:] &&\models \varphi_2 \land \forall i, 0\leq i < j, \ M, \rho[i:] \models \varphi_1, \nonumber
	\end{alignat}
where $\rho[i:]$ be the $i$-th suffix $\rho[i:]=s_ia_is_{i+1} \ldots $.

The next operator {\bf X} requires that $\varphi$ is satisfied by the next state suffix of $\rho$. The until operator {\bf U} requires that $\varphi_1$ holds true until $\varphi_2$ becomes true over the path $\rho$.
In the following, we write $ \rho \models \varphi $ for simplicity without referring to $ M $.

For any policy $\pi$, the probability of all paths starting from $s_{init}$ on the MDP $M$ that satisfy an LTL formula $\varphi$ under the policy $\pi$, or the satisfaction probability under $\pi$ is defined as
\begin{align*}
Pr^{M}_{\pi}(s_{init} \! \models \varphi) := Pr^{M}_{\pi}(\{ \rho_{init} \! \in \! InfPath^{M}_{\pi} ; \rho_{init} \! \models \varphi \}).
\end{align*}
We say that an LTL formula $\varphi$ is positively satisfied by a positional policy $\pi$ if
\begin{align*}
Pr^{M}_{\pi}(s_{init} \models \varphi) > 0.
\end{align*}

\label{def5}
\end{definition}

Any LTL formula $\varphi$ can be converted into various $\omega$-automata, namely finite state machines that recognize 
all infinite words satisfying $\varphi$.
 We review a generalized B\"{u}chi automaton at the beginning, and then introduce a limit-deterministic generalized B\"{u}chi automaton \cite{HKAKPL2019}.

\begin{definition}
  A transition-based generalized B\"{u}chi automaton (tGBA) is a tuple $B = (X,\ x_{init},\ \Sigma,\ \delta,\ \mathcal{F})$, where $X$ is a finite set of states, $x_{init} \in X$ is the initial state, $\Sigma$ is an input alphabet including $\varepsilon$, $\delta \subset  X\times \Sigma \times X$ is a set of transitions, and $\mathcal{F} = \{F_1,\ldots,F_n\}$ is an acceptance condition, where for each $ j \in \{1,\ldots,n\}$, $F_j \subset \delta$ is a set of accepting transitions and called an accepting set. We refer to a tGBA with one accepting set as a tBA.

  An infinite sequence $r = x_0\sigma_0x_1 \ldots \in X (\Sigma X)^{\omega}$ is called an infinite run if $(x_i, \sigma_{i}, x_{i+1}) \in \delta\ $ for any $ i\in \mathbb{N}_0$. An infinite word $w = \sigma_0\sigma_1 \ldots \in \Sigma^{\omega}$ is accepted by $B_{\varphi}$ if and only if there exists an infinite run $r = x_0 \sigma_0 x_1 \ldots$ starting from $x_0 = x_{init}$ such that $inf(r) \cap F_j \neq \emptyset\ $ for each $F_j \in \mathcal{F}$, where $inf(r)$ is the set of transitions that occur infinitely often in the run $r$.
\end{definition}

\begin{definition}
  A transition-based limit-deterministic generalized B\"{u}chi automaton (tLDGBA) is a tGBA $B = (X, x_{init},\Sigma,\delta,\mathcal{F})$ such that $X$ is partitioned into two disjoint sets $X_{initial}$ and $X_{final}$ such that
  \begin{itemize}
    \item $F_j \subset X_{final} \times \Sigma \times X_{final}$, $\forall j \in \{ 1,...,n \}$,
    \item $| \{ (x, \sigma, x^{\prime}) \in \delta; \sigma \! \in \! \Sigma, x^{\prime} \in X_{final} \} | \! = \! 1$, $\forall x \! \in \! X_{final}$,
    \item $| \{ (x, \sigma, x^{\prime}) \in \delta; \sigma \! \in \! \Sigma, x^{\prime} \in X_{initial} \} |$=0, $\forall x \! \in \! X_{final}$,
    \item 
    $ \forall (x, \varepsilon, x') \in \delta, \ x \in X_{initial} \land x' \in X_{final} $.
\end{itemize}
\end{definition}
  An $\varepsilon$-transition enables the tLDGBA to change its state with no input
  and reflects a single ``guess" from $X_{initial}$ to $X_{final}$. Note that by the construction in \cite{SEJK2016}, the transitions in each part are deterministic except for $\varepsilon$-transitions from $X_{intial}$ to $X_{final}$.
It is known that, for any LTL formula $ \varphi $, there exists a tLDGBA that accepts all words satisfying $\varphi$ \cite{SEJK2016}. We refer to a tLDGBA with one accepting set as a tLDBA.
In particular, we represent a tLDGBA recognizing an LTL formula $\varphi$ as $B_{\varphi}$, whose input alphabet is given by $ \Sigma = 2^{AP} \cup \{ \varepsilon \} $.

\section{Reinforcement-Learning-Based Synthesis of Control Policy}
We introduce an automaton augmented with binary-valued vectors. The automaton can 
ensure that transitions in each accepting set occur infinitely often.

Let $V = \{ (v_1, \ldots ,v_n)^T\ ;\ v_i \in \{ 0,1 \},\ i \in \{ 1, \ldots ,n \} \}$ be a set of binary-valued vectors, and let $\bm{1}$ and $\bm{0}$ be the $n$-dimentional vectors with all elements 1 and 0, respectively.
In order to augment a tLDGBA $B_{\varphi}$, we introduce three functions $visitf:\delta \rightarrow V$, $reset:V \rightarrow V$, and $Max:V\times V \rightarrow V$ as follows.
For any $e \in \delta$, $visitf(e) = (v_1, \ldots ,v_n)^T$, where $ v_i = 1 $ if $ e \in F_i $ and $ v_i=0 $ otherwise.
For any $v \in V$, $ reset(v) = \bm{0} $ if $ v = \bm{1} $ and $ reset(v) = v $ otherwise.
For any $v,u \in V$, $Max(v,u) = (l_1,\ldots ,l_n)^T$, where $l_i = max\{v_i, u_i\} $ for any $i\in \{1, \ldots ,n\}$.

Each vector $v$ is called a memory vector and represents which accepting sets have been visited. The function $visitf$ returns a binary vector whose $i$-th element is 1 if and only if a transition in the accepting set $F_i$ occurs. The function $reset$ returns the zero vector $\bm{0}$ if at least one transition in each accepting set has occurred after the latest reset. Otherwise, it returns the input vector without change.

\begin{definition}
   For a tLDGBA $B_{\varphi} = (X,x_{init},\Sigma,\delta,\mathcal{F})$, its augmented automaton is a tLDGBA $\bar{B}_{\varphi} = (\bar{X},\bar{x}_{init},$ $\bar{\Sigma},\bar{\delta},\bar{\mathcal{F}})$, where 
   $\bar{X} = X\times V$, 
   $\bar{x}_{init} = (x_{init}, \bm{0})$, 
   $\bar{\Sigma} = \Sigma$, 
   $\bar{\delta}$ is defined as $\bar{\delta}$ = $\{ ((x,v), \bar{\sigma}, (x^{\prime},v^{\prime})) \in \bar{X} \times \bar{\Sigma} \times \bar{X} ;\ (x,\bar{\sigma},x^{\prime}) \in \delta, v^{\prime} = reset(Max(v,visitf((x,\bar{\sigma},x^{\prime})))) \}$, and
   $\mathcal{\bar{F}} = \{ \bar{F_1},$ $ \ldots ,\bar{F_n} \}$ is defined as $\bar{F_j} = \{ ((x,v), \bar{\sigma}, (x^{\prime},v^{\prime})) \in \bar{\delta}\ ;(x, \bar{\sigma}, x^{\prime}) \in F_j,$ $ v_j = 0 \}$ for each $ j \in \{1,...,n\}$.
   \label{augment_def}
\end{definition}

The augmented tLDGBA $\bar{B}_{\varphi}$ keeps track of previous visits to the accepting sets of $B_{\varphi}$. 
Intuitively, 
along a run of $\bar{B}_\varphi$, 
a memory vector $v$ is reset to $\bm{0}$ when at least one transition in each accepting set of the original tLDGBA $B_{\varphi}$ has occurred. 
We now show that a tLDGBA and its augmented automaton accept the same language.
Let $ \mathcal{L}(B) \subseteq \Sigma^\omega $ be the accepted language of an automaton $ B $ with the alphabet $ \Sigma $, namely the set of all infinite words accepted by $ B $.

\begin{prop}
Let $ B = (X,x_{init},\Sigma,\delta,\mathcal{F}) $ and $ \bar{B} = (\bar{X},\bar{x}_{init},\bar{\Sigma},\bar{\delta},\bar{\mathcal{F}}) $ be an arbitrary tLDGBA and its augmentation, respectively.
Then, we have $ \mathcal{L}(B) = \mathcal{L}(\bar{B}) $.
\end{prop}

\begin{IEEEproof}
Recall that $ \Sigma = \bar{\Sigma} $.
We prove set inclusions in both directions.
\begin{itemize}
	\item[$ \subset $:]
	Consider any $ w = \sigma_0 \sigma_1 \ldots \in \mathcal{L}(B) $.
	Then, there exists a run $ r = x_0 \sigma_0 x_1 $ $\sigma_1 x_2 \ldots \in X(\Sigma X)^\omega $ of $ B $ such that $ x_0 = x_{init} $ and $ inf(r) \cap F_j \neq \emptyset $ for each $ F_j \in \mathcal{F} $.
	For the run $ r $, we construct a sequence $ \bar{r} = \bar{x}_0 \bar{\sigma}_0 \bar{x}_1 \bar{\sigma}_1 \bar{x}_2 \ldots \in \bar{X}(\bar{\Sigma} \bar{X})^\omega $ satisfying $ \bar{x}_i = (x_i, {v}_i) $ and $ \bar{\sigma}_i = \sigma_i $ for any $ i \in \mathbb{N} $, where $ {v}_0 = \mathbf{0} $ and 
	\begin{align*}
	     {v}_{i+1} \!=\! reset \Big( \! Max \big( {v}_i, visitf((x_i,\bar{\sigma}_i,x_{i+1})) \big) \! \Big), \forall i \in \mathbb{N}.
	\end{align*}
	Clearly from the construction, we have $ (\bar{x}_i, \bar{\sigma}_i, \bar{x}_{i+1}) \in \bar{\delta} $ for any $ i \in \mathbb{N} $.
	Thus, $ \bar{r} $ is a run of $ \bar{B} $ starting from $ \bar{x}_0 = (x_{init}, \mathbf{0}) = \bar{x}_{init} $.

	We now show that $ inf(\bar{r}) \cap \bar{F}_j \neq \emptyset $ for each $ \bar{F}_j \in \bar{\mathcal{F}} $.
	Since $ inf(r) \cap F_j \neq \emptyset $ for each $ F_j \in \mathcal{F} $, we have
	\begin{align*}
	    & inf\!(\bar{r}) \!\cap\! \{ \!( \bar{x}, \!\bar{\sigma}, \!\bar{x}') \!\in \!\bar{\delta}; 
	    visitf\!(\! ( [\![\bar{x}]\!]_X, \!\bar{\sigma},\! [\![\bar{x}']\!]_X\!)\!)_j \!\!=\! 1 \} 
	    \!\neq\! \emptyset
	\end{align*}
	for any $ j \in \{ 1, \ldots, n \} $, 
	where $ [\![ (x,v) ]\!]_X = x $ for each $ (x,v) \in \bar{X} $.
	By the construction of $ \bar{r} $, therefore, there are infinitely many indices $ l \in \mathbb{N} $ with $ v_l = \mathbf{0} $.
	Let $ l_1, l_2 \in \mathbb{N} $ be arbitrary nonnegative integers such that $ l_1 < l_2 $, $ v_{l_1} = v_{l_2} = \mathbf{0} $, and $ v_{l'} \neq \mathbf{0} $ for any $ l' \in \{ l_1+1, \ldots, l_2-1\} $.
	Then, 
	\begin{align*}
    	\forall j \in \{1, \ldots, n\}, \ & \exists k \in \{  l_1, l_1+1, \ldots, l_2-1 \}, \\
    	& (x_k, \sigma_k, x_{k+1}) \in F_j \land (v_k)_j = 0 ,
	\end{align*}
	where $ (v_k)_j $ is the $ j $-th element of $ v_k $.
	Hence, we have $ inf(\bar{r}) \cap \bar{F}_j \neq \emptyset $ for each $ \bar{F}_j \in \bar{\mathcal{F}} $, which implies $ w \in \mathcal{L}(\bar{B}) $.

	\item[$ \supset $:]
	Consider any $ \bar{w} \in \bar{\sigma}_0 \bar{\sigma}_1 \ldots \in \mathcal{L}(\bar{B}) $.
	Then, there exists a run $ \bar{r} = \bar{x}_0 \bar{\sigma}_0 \bar{x}_1 \bar{\sigma}_1 $ $ \bar{x}_2 \ldots \in \bar{X}(\bar{\Sigma} \bar{X})^\omega $ of $ \bar{B} $ such that $ \bar{x}_0 = \bar{x}_{init} $ and $ inf(\bar{r}) \cap \bar{F}_j \neq \emptyset $ for each $ \bar{F}_j \in \bar{\mathcal{F}} $, i.e.,
	\begin{align}
	    \label{eq:inf_r}
	\forall j \in & \{1, \ldots, n\}, \
	 \forall k \in \mathbb{N}, \ \exists l \geq k, \nonumber \\
	& ( [\![ \bar{x}_l ]\!]_X , \bar{\sigma}_l, [\![ \bar{x}_{l+1} ]\!]_X) \in F_j \land (\bar{v}_l)_j = 0 .
	\end{align}
	For the run $ \bar{r} $, we construct a sequence $ r = x_0 \sigma_0 x_1 \sigma_1 x_2 \ldots \in X(\Sigma X)^\omega $ such that $ x_i = [\![ \bar{x}_i ]\!]_X $ and $ \sigma_i = \bar{\sigma}_i $ for any $ i \in \mathbb{N} $.
	It is clear that $ r $ is a run of $ B $ starting from $ x_0 = x_{init} $.
	It holds by Eq.~\eqref{eq:inf_r} that $ inf(r) \cap F_j \neq \emptyset $ for each $ F_j \in \mathcal{F} $, which implies $ \bar{w} \in \mathcal{L}(B) $.
\end{itemize}
\end{IEEEproof}

\begin{figure}[htbp]
   \centering
   \includegraphics[bb=0 0 247 80,scale=0.75]{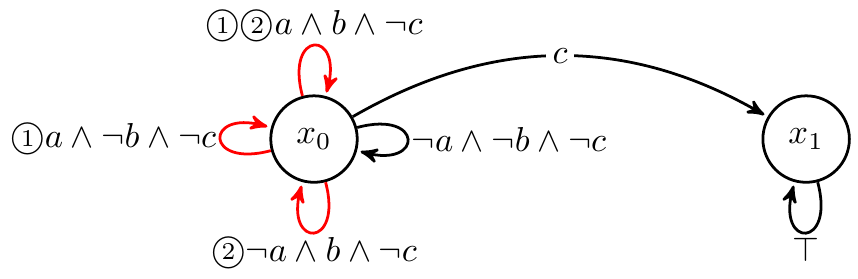}
   \caption{The tLDGBA recognizing the LTL formula $\text{{\bf GF}}a \wedge \text{{\bf GF}}b \wedge \text{{\bf G}}\neg c$, where $ X = \{x_0, x_1\} = X_{final} $ and $x_{init} = x_0$. Red arcs are accepting transitions that are numbered in accordance with the accepting sets they belong to.}
   \label{automaton}
\end{figure}

\begin{figure}[htbp]
   \centering
   \includegraphics[bb=0 0 326 207,scale=0.75]{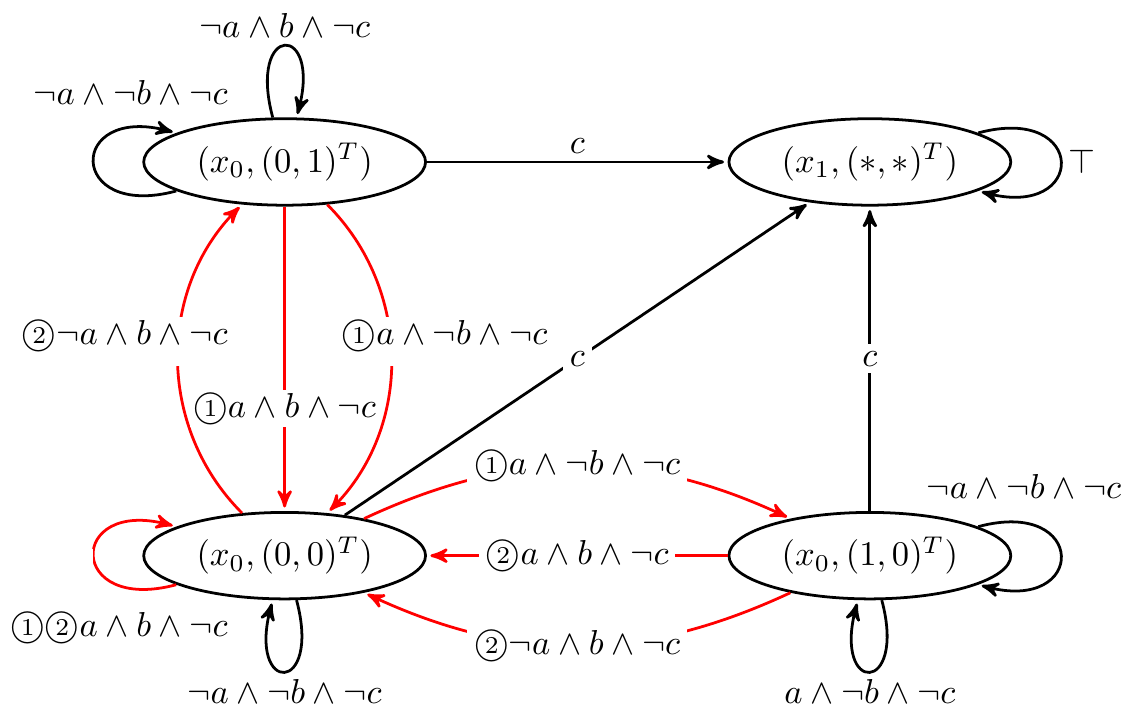}
   \caption{The augmented automaton for the tLDGBA in Fig.~\ref{automaton} recognizing the LTL formula $\text{{\bf GF}}a \wedge \text{{\bf GF}}b \wedge \text{{\bf G}}\neg c$, 
   where the initial state is $(x_0, (0,0)^T )$. Red arcs are accepting transitions that are numbered in accordance with the accepting sets they belong to. All states corresponding to $x_1$ are merged into $(x_1, (*,*)^T )$.}
   \label{automaton_aug}
\end{figure}

For example, shown in Figs.\ \ref{automaton} and \ref{automaton_aug} are a tLDGBA and its augmented automaton, respectively, associated with the following LTL formula.
\begin{align}
  \varphi = \text{{\bf GF}}a \land \text{{\bf GF}}b \land \text{{\bf G}}\neg c.
  \label{ltl}
\end{align}
The acceptance condition ${\mathcal F}$ of the tLDGBA is given by ${\mathcal F} =$ $ \{ F_1,F_2 \}$, where $F_1=\{ (x_0, \{ a \}, x_0),\ (x_0, \{ a,b \}, x_0) \}$ and $F_2 = \{ (x_0, \{ b \}, x_0),\ (x_0, \{ a,b \}, x_0) \}$. 
Practically, states in a strongly connected component that contains no accepting transitions can be merged as shown in Fig.\ \ref{automaton_aug}.


We modify the standard definition of a product MDP to deal with $\varepsilon$-transitions in the augmented automaton.
\begin{definition}
  Given an augmented tLDGBA $\bar{B}_{\varphi}$ and an MDP $M$, a tuple $M \otimes \bar{B}_{\varphi} = M^{\otimes} = (S^{\otimes}, A^{\otimes},s_{init}^{\otimes}, P^{\otimes}, \delta^{\otimes},$ $ {\mathcal F}^{\otimes})$ is a product MDP, where
  $S^{\otimes} = S \times \bar{X}$ is the finite set of states;
  $A^{\otimes}$  is the finite set of actions such that $A^{\otimes}=A \cup \{ \varepsilon_{\bar{x}^{\prime}} ; \exists \bar{x}^{\prime}\! \in \! X \text{ s.t. } (\bar{x},\varepsilon,\bar{x}^{\prime}) \in \bar{\delta} \}$, where $\varepsilon_{\bar{x}^{\prime}}$ is the action for the $\varepsilon$-transition to the state $\bar{x}^{\prime}\! \in\! \bar{X}$;
  $s_{init}^{\otimes} = (s_{init},\bar{x}_{init})$ is the initial state; $P^{\otimes} : S^{\otimes} \times S^{\otimes} \times A^{\otimes} \rightarrow [0,1]$ is the transition probability function defined as 
  \begin{align*}
    &P^{\otimes}(s^{\otimes \prime} | s^{\otimes}, a) \\ &=
    \left\{
    \begin{aligned}
      &P(s^{\prime} | s, a) &   &\text{if}\  (\bar{x}, L((s,a,s^{\prime})), \bar{x}^{\prime}) \in \bar{\delta}, a \in \mathcal{A}(s)\\
      &1 &   &\text{if}\ s=s^{\prime}, (\bar{x}, \varepsilon, \bar{x}^{\prime}) \in  \bar{\delta}, a=\varepsilon_{\bar{x}^{\prime}},\\
      &0 &   &\text{otherwise} ,
    \end{aligned}
    \right. \nonumber
  \end{align*}
  where $s^{\otimes}=(s,(x,v))$ and $s^{\otimes \prime}=(s^{\prime},(x^{\prime},v^{\prime}))$;
  $\delta^{\otimes} = \{ (s^{\otimes}, a, s^{\otimes \prime})\in S^{\otimes} \times A^{\otimes} \times S^{\otimes} ; P^{\otimes}(s^{\otimes \prime} | s^{\otimes}, a) > 0 \}$ is the set of transitions;
  and ${\mathcal F}^{\otimes} = \{ \bar{F}^{\otimes}_1, \ldots ,\bar{F}^{\otimes}_n \}$ is the acceptance condition, where $\bar{F}^{\otimes}_i = \{ ((s,\bar{x}), a, (s^{\prime}, \bar{x}^{\prime})) \in \delta^{\otimes} ; (\bar{x}, L(s,a,s^{\prime}), \bar{x}^{\prime}) \in \bar{F}_i \}$ for each $ i \in \{ 1, \ldots ,n \}$.

\label{def9}
\end{definition}

\begin{definition}
  The reward function $\mathcal{R} :S^{\otimes} \times A^{\otimes} \times S^{\otimes} \rightarrow {\mathbb R}_{\geq 0}$ is defined as
  \begin{align}
    \mathcal{R}(s^{\otimes}, a, s^{\otimes \prime}) =
    \left\{
    \begin{aligned}
      &r_p \  \text{if}\ \exists i \in \! \{ 1, \ldots ,n \},\ (s^{\otimes}, a, s^{\otimes \prime}) \in \bar{F}^{\otimes}_i \!,\\
      &0   \ \ \text{otherwise},
    \end{aligned}
    \right. \nonumber
  \end{align}
  where $r_p$ is a positive value.
  \label{def10}
\end{definition}

\begin{remark}
  When constructing a tBA from a tGBA, the order of visits to accepting sets of the tGBA is fixed. Consequently, the rewards based on the acceptance condition of the tBA tends to be sparse. 
  The sparsity is critical in RL-based controller synthesis problems. 
  The augmentation of the tGBA relaxes the sparsity since the augmented tGBA has all of the order of visits to all accepting sets of the original tGBA. 
  For the acceptance condition $\mathcal{F}$ of the tGBA, the size of the state space of the augmented tGBA is about $\frac{2^{|\mathcal{F}|}-1}{|\mathcal{F}|}$ times larger than the tBA constructed from the tGBA. 
  However, the number of accepting transitions to all transitions in the augmented tGBA is much greater than the tBA. 
  Therefore, our proposed method is expected to be better than using the tLDBA in the sense of sample efficiency.
\end{remark}

In the following, we show that a positional policy positively satisfying $\varphi$ on the product MDP $M^{\otimes}$ is synthesized by using the reward function $\mathcal{R}$ in Definition \ref{def10}, which is based on the acceptance condition of $ M^\otimes $. 

For a Markov chain $MC^{\otimes}_{\pi}$ induced by a product MDP $M^{\otimes}$ with a positional policy $\pi$, let $S^{\otimes}_{\pi}= T^{\otimes}_{\pi} \cup R^{\otimes 1}_{\pi} \cup \ldots \cup R^{\otimes h}_{\pi}$ be the set of states in $MC^{\otimes}_{\pi}$, where $T^{\otimes}_{\pi}$ is the set of transient states and $R^{\otimes i}_{\pi}$ is the recurrent class for each $i \in \{ 1, \ldots ,h \}$, and let $R(MC^{\otimes}_{\pi})$ be the set of all recurrent states in $MC^{\otimes}_{\pi}$. Let $\delta^{\otimes i}_{\pi}$ be the set of transitions in a recurrent class $R^{\otimes i}_{\pi}$, namely $\delta^{\otimes i}_{\pi} = \{ (s^{\otimes},a,s^{\otimes \prime}) \in \delta^{\otimes} ; s^{\otimes} \in R^{\otimes i}_{\pi},\ P^{\otimes}(s^{\otimes \prime}|s^{\otimes},a) > 0 \}$, and let $P^{\otimes}_{\pi}$ : $S^{\otimes}_{\pi} \times S^{\otimes}_{\pi} \rightarrow [0,1]$ be the transition probability function under $\pi$.

\begin{lemma}
  For any policy $\pi$ and any recurrent class $R^{\otimes i}_{\pi}$ in the Markov chain $MC^{\otimes}_{\pi}$,
  $MC^{\otimes}_{\pi}$ satisfies one of the following conditions.
  \vspace{2mm}
  \begin{enumerate}
    \item $\delta^{\otimes i}_{\pi} \cap \bar{F}^{\otimes}_j \neq \emptyset\ $, $ \forall j \in \{ 1, \ldots ,n \}$,
    \item $\delta^{\otimes i}_{\pi} \cap \bar{F}^{\otimes}_j = \emptyset\ $, $ \forall j \in \{ 1, \ldots ,n \}$.
  \end{enumerate}
  \label{lemma1}
\end{lemma}

\begin{IEEEproof}
  Suppose that $MC^{\otimes}_{\pi}$ satisfies neither conditions 1 nor 2. Then, there exist a policy $\pi$, $i \in \{ 1, \ldots ,h \}$, and $j_1$, $j_2$ $\in \{ 1, \ldots ,n \}$ such that $\delta^{\otimes i}_{\pi} \cap \bar{F}^{\otimes}_{j_1} = \emptyset$ and $\delta^{\otimes i}_{\pi} \cap \bar{F}^{\otimes}_{j_2} \neq \emptyset$. In other words, there exists a nonempty and proper subset $J \in 2^{\{ 1, \ldots ,n \}} \setminus \{ \{ 1, \ldots ,n \}, \emptyset \}$ such that $ \delta^{\otimes i}_{\pi} \cap \bar{F}^{\otimes}_j \neq \emptyset $ for any $j \in J$.
   For any transition $ (s,a,s^{\prime}) \in \delta^{\otimes i}_{\pi} \cap \bar{F}^{\otimes}_j$, the following equation holds by the properties of the recurrent states in $MC^{\otimes}_{\pi}$\cite{ESS}.
  \begin{align}
    \sum_{k=0}^{\infty} p^k((s,a,s^{\prime}),(s,a,s^{\prime})) = \infty,
    \label{eq15}
  \end{align}
  where $p^k((s,a,s^{\prime}),(s,a,s^{ \prime}))$ is the probability that the transition $(s,a,s^{\prime})$ reoccurs after it occurs in $k$ time steps. 
  Eq. (\ref{eq15}) means that each transition in $R^{\otimes i}_{\pi}$ occurs infinitely often. However, the memory vector $v$ is never reset in $R^{\otimes i}_{\pi}$ by the assumption. This directly contradicts Eq.\ (\ref{eq15}).
\end{IEEEproof}

Lemma \ref{lemma1} implies that, for an LTL formula $\varphi$, if a policy $\pi$ does not satisfy $\varphi$, then the agent obtains no reward in recurrent classes; otherwise there exists at least one recurrent class where the agent obtains rewards infinitely often.

\begin{theorem}
  For a product MDP $M^{\otimes}$ of an MDP $M$ and an augmented tLDGBA $\bar{B}_{\varphi}$ corresponding to a given LTL formula $\varphi$ and the reward function given by Definition \ref{def10}, if there exists a positional policy positively satisfying $\varphi$ on $M^{\otimes}$, then there exists a discount factor $\gamma^{\ast}$ such that any algorithm that maximizes the expected discounted reward with $\gamma > \gamma^{\ast}$ will find a positional policy positively satisfying $\varphi$ on $M^{\otimes}$.
  \label{theorem1}
\end{theorem}

\begin{IEEEproof}
  Suppose that $\pi^{\ast}$ is an optimal policy but does not satisfy the LTL formula $\varphi$. Then, for any recurrent class $R^{\otimes i}_{{\pi}^{\ast}}$ in the Markov chain $MC^{\otimes}_{{\pi}^{\ast}}$ and any accepting set $\bar{F}^{\otimes}_j$ of the product MDP $M^{\otimes}$,  $\delta^{\otimes i}_{\pi^{\ast}} \cap \bar{F}^{\otimes}_j = \emptyset$
  holds by Lemma \ref{lemma1}. Thus, the agent under the policy $\pi^{\ast}$ can obtain rewards only in the set of transient states. We consider the best scenario in the assumption. Let $p^k(s,s^{\prime})$ be the probability of going to a state $s^{\prime}$ in $k$ time steps after leaving the state $s$, and let $Post(T^{\otimes}_{\pi})$ be the set of states in recurrent classes that can be transitioned from states in $T^{\otimes}_{\pi}$ by one action. For the initial state $s_{init}$ in the set of transient states, it holds that
  \begin{align*}
    &V^{\pi^{\ast}}\!(s_{init}) \\
     = & \sum_{k=0}^{\infty} \sum_{s \in T^{\otimes}_{\pi^{\ast}}} \!\!\gamma^k p^k(s_{init}, s) 
      \!\!\!\!\sum_{s^{\prime} \in T^{\otimes}_{\pi^{\ast}} \cup Post(T^{\otimes}_{\pi^{\ast}})} \!\!\!\!\!\!P^{\otimes}_{\pi^{\ast}}\!(s^{\prime}| s) \mathcal{R}(s, a, s^{\prime})\nonumber \\
     \leq & r_p \sum_{k=0}^{\infty} \sum_{s \in T^{\otimes}_{\pi^{\ast}}} \gamma^k p^k(s_{init}, s), \nonumber
  \label{eqth11}
  \end{align*}
  where the action $a$ is selected by $\pi^{\ast}$. By the property of the transient states, for any state $s^{\otimes}$ in $T^{\otimes}_{\pi^{\ast}}$, there exists a bounded positive value $m$ such that $ \sum_{k=0}^{\infty} \gamma^k p^k(s_{init}, s) \leq \sum_{k=0}^{\infty} p^k(s_{init}, s) < m$ \cite{ESS}. Therefore, there exists a bounded positive value $\bar{m}$ such that $V^{\pi^{\ast}}(s_{init}) < \bar{m}$.
  Let $\bar{\pi}$ be a positional policy satisfying $\varphi$. We consider the following two cases.
  \begin{enumerate}
    \vspace{2mm}
    \item Assume that the initial state $s_{init}$ is in a recurrent class $R^{\otimes i}_{\bar{\pi}}$ for some $ i \in \{1,\ldots,h\} $.
    For any accepting set $\bar{F}^{\otimes}_j$, $\delta^{\otimes i}_{\bar{\pi}} \cap \bar{F}^{\otimes}_j \neq \emptyset$ holds by the definition of $\bar{\pi}$. The expected discounted reward for $s_{init}$ is given by
    \begin{align*}
      & V^{\bar{\pi}}(s_{init}) \\ 
       = &\sum_{k=0}^{\infty} \sum_{s \in R^{\otimes i}_{\bar{\pi}}} \! \gamma^k p^k(s_{init}, s) \!\! 
       \sum_{s^{\prime} \in R^{\otimes i}_{\bar{\pi}}}  P^{\otimes}_{\bar{\pi}}(s^{\prime} | s) \mathcal{R}(s, a, s^{\prime}), \nonumber
    \end{align*}
    where the action $a$ is selected by $\bar{\pi}$. Since $s_{init}$ is in $R^{\otimes i}_{\bar{\pi}}$, there exists a positive number $\bar{k} = \min \{ k\ ;\ k \geq n, p^{k}(s_{init}, s_{init}) > 0 \}$ \cite{ESS}. We consider the worst scenario in this case. It holds that
    \begin{align}
      V^{\bar{\pi}}(s_{init})
       \geq & \sum_{k=n}^{\infty} p^{k}(s_{init}, s_{init}) \sum_{i=1}^{n} \gamma^{k-i} r_p \nonumber \\
       \geq & \sum_{k=1}^{\infty} p^{k \bar{k}}(s_{init}, s_{init}) \sum_{i=0}^{n-1} \gamma^{k \bar{k} - i} r_p \nonumber \\
       > & r_p \sum_{k=1}^{\infty} \gamma^{k \bar{k}} p^{k \bar{k}}(s_{init}, s_{init}), \nonumber
    \end{align}
whereas all states in $R(MC^{\otimes}_{\bar{\pi}})$ are positive recurrent because $|S^{\otimes}| < \infty$ \cite{ISP}. Obviously, $p^{k \bar{k}}(s_{init}, s_{init}) \geq (p^{\bar{k}}(s_{init}, s_{init}))^k > 0$ holds for any $k \in (0, \infty)$ by the Chapman-Kolmogorov equation \cite{ESS}. Furthermore, we have $\lim_{k \rightarrow \infty} p^{k \bar{k}}(s_{init}, s_{init}) > 0$ by the property of irreducibility and positive recurrence \cite{SM}. Hence, there exists $\bar{p}$ such that $0<\bar{p}<p^{k \bar{k}}(s_{init}, s_{init})$ for any $k \in (0, \infty]$ and we have
    \begin{align}
       V^{\bar{\pi}}(s_{init}) > & r_p \bar{p} \frac{\gamma^{\bar{k}}}{ 1 - \gamma^{\bar{k}} }. \nonumber
    \end{align}

    Therefore, for any $r_p < \infty$ and any $\bar{m} \in (V^{\pi^{\ast}}(s_{init}), \infty)$, there exists $\gamma^{\ast}<1$ such that $\gamma > \gamma^{\ast}$ implies $V^{\bar{\pi}}(s_{init}) > r_p \bar{p} \frac{\gamma^{\bar{k}}}{ 1 - \gamma^{\bar{k}} } > \bar{m}.$

    \item Assume that the initial state $s_{init}$ is in the set of transient states $T_{\bar{\pi}}^{\otimes}$.
    $P^{M^{\otimes}}_{\bar{\pi}}(s_{init} \models \varphi) > 0$ holds by the definition of $\bar{\pi}$. For a recurrent class $R^{\otimes i}_{\bar{\pi}}$ such that $\delta^{\otimes i}_{\bar{\pi}} \cap \bar{F}^{\otimes}_j \neq \emptyset$ for each accepting set
    $\bar{F}^{\otimes}_j$, there exist a number $\bar{l} > 0$, a state $\hat{s}$ in $Post(T^{\otimes}_{\bar{\pi}}) \cap R^{\otimes i}_{\bar{\pi}}$, and a subset of transient states $\{ s_1, \ldots , s_{\bar{l}-1} \} \subset T^\otimes_{\bar{\pi}}$ such that $p(s_{init}, s_1)>0$, $p(s_{i}, s_{i+1})>0$ for $i \in \{ 1,\ldots, $ $\bar{l}-2 \}$, and $p(s_{\bar{l}-1}, \hat{s})>0$ by the property of transient states.
    Hence, it holds that $p^{\bar{l}}(s_{init}, \hat{s}) > 0$ for the state $\hat{s}$. Thus, by ignoring rewards in $T^{\otimes}_{\bar{\pi}}$, we have
     \begin{align}
        V^{\bar{\pi}}(s_{init}) 
        \geq\ & \gamma^{\bar{l}} p^{\bar{l}}(s_{init}, \hat{s}) \sum_{k=0}^{\infty} \sum_{s^{\prime} \in R^{\otimes i}_{\bar{\pi}}} \gamma^k p^k(\hat{s}, s^{\prime}) \nonumber \\
        & \sum_{s^{\prime \prime} \in R^{\otimes i}_{\bar{\pi}}} P^{\otimes}_{\bar{\pi}}(s^{\prime \prime} | s^{\prime}) \mathcal{R}(s^{\prime}, a, s^{\prime \prime}) \nonumber \\
        >\ & \gamma^{\bar{l}} p^{\bar{l}}(s_{init}, \hat{s}) r_p \bar{p} \frac{\gamma^{\bar{k}^{\prime}}}{ 1 - \gamma^{\bar{k}^{\prime}} }, \nonumber
     \end{align}
     where $\bar{k}^{\prime}  \geq n$ is a constant and $0<\bar{p}< p^{k \bar{k}^{\prime}}(\hat{s}, \hat{s})$ for any $k \in (0, \infty]$.
     Therefore, for any $r_p < \infty$ and any $\bar{m} \in (V^{\pi^{\ast}}(s_{init}), \infty)$, there exists $\gamma^{\ast}<1$ such that $\gamma > \gamma^{\ast}$ implies
     $V^{\bar{\pi}}(s_{init}) > \gamma^{\bar{l}} p^{\bar{l}}(s_{init}, \hat{s}) \frac{r_p \bar{p} \gamma^{\bar{k}^{\prime}}}{ 1 - \gamma^{\bar{k}^{\prime}} } > \bar{m}$.
  \end{enumerate}

The results contradict the optimality assumption of $\pi^{\ast}$.
\end{IEEEproof}

Theorem \ref{theorem1} implies that, for the product MDP $M^{\otimes}$ of an MDP $M$ and an augmented tLDGBA corresponding to a given LTL formula $\varphi$, 
we obtain a positional policy positively satisfying $\varphi$ on $M^{\otimes}$ by an algorithm maximizing the expected discounted reward with a discount factor sufficiently close to 1 if there exists a positional policy on $M^{\otimes}$ satisfying $\varphi$.

\section{Example}

\begin{figure}[tbp]
    \centering
    \includegraphics[bb=0 0 377 290,height=3.5cm,width=5cm]{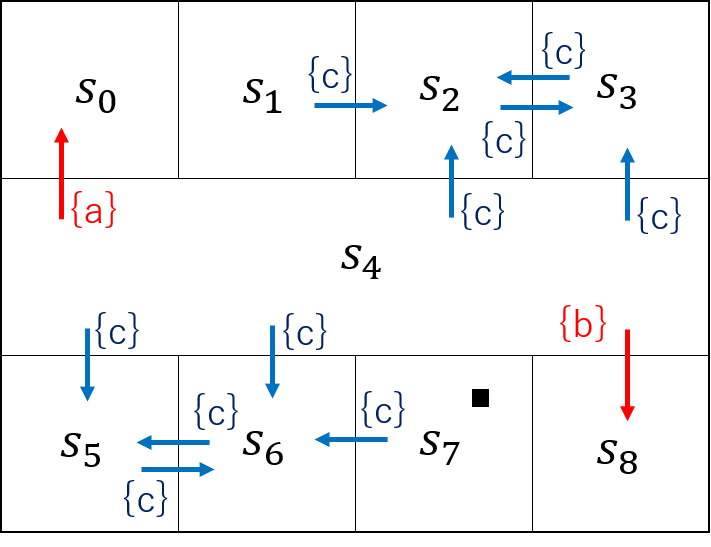}
    \caption{The environment consisting of eight rooms and one corridor. Red arcs are the transitions that we want to occur infinitely often, while blue arcs are the transitions that we never want to occur. $s_7$ is the initial state.}
    \label{Grid1}
\end{figure}

In this section, 
we apply the proposed method to a path planning problem of a robot in an environment consisting of eight rooms and one corridor as shown in Fig.\ \ref{Grid1}. The state $s_7$ is the initial state and the action space is specified with $\mathcal{A}(s) = \{ Right, Left, Up, Down \}$ for any state $s \neq s_4$ and $\mathcal{A}(s_4) = \{ to\_s_0, to\_s_1, to\_s_2, to\_s_3, to\_s_5, $ $to\_s_6, to\_s_7, to\_s_8 \}$, where $to\_s_i$ means attempting to go to the state $s_i$ for $i \in \{0,\ 1,\ 2,\ 3,\ 5,\ 6,\ 7,\ 8 \}$. The robot moves in the intended direction with probability 0.9 and it stays in the same state with probability 0.1 if it is in the state $s_4$. In the states other than $s_4$, it moves in the intended direction with probability 0.9 and it moves in the opposite direction to that it intended to go with probability 0.1. If the robot tries to go to outside the environment, it stays in the same state. The labeling function is as follows.
\begin{align*}
      & L((s, act, s^{\prime})) =
      \left\{
      \begin{aligned}
        & \{ c \} &  & \text{if }s^{\prime} = s_i,\ i \in \{ 2,3,5,6 \}, \nonumber \\
        & \{ a \} &  & \text{if }(s,act,s^{\prime})=(s_4,to\_s_0,s_0), \nonumber \\
        & \{ b \} &  & \text{if }(s,act,s^{\prime})=(s_4,to\_s_8, s_8), \nonumber \\
        & \emptyset &  & \text{otherwise}.
      \end{aligned}
      \right.
    \end{align*}

In the example, the robot tries to take two transitions that we want to occur infinitely often, represented by arcs labeled by \{$a$\} and \{$b$\}, while avoiding unsafe transitions represented by the arcs labeled by \{{\it c}\}. This is formally specified by the LTL formula given by Eq.~(\ref{ltl}).
The LTL formula requires the robot to keep on entering the two rooms $s_0$ and $s_8$ from the corridor $s_4$ regardless of the order of entries, while avoiding entering the four rooms $s_2$, $s_3$, $s_5$, and $s_6$. 
The tLDGBA $B_{\varphi}$ and its augmented automaton $\bar{B}_{\varphi} $ are shown in Figs.\ \ref{automaton} and \ref{automaton_aug}, respectively.

Through the above scenario, 
we compare our approach with 1) a case where we first convert the tLDGBA into a tLDBA, for which the augmentation makes no change, and thus a reward function in Definition \ref{def10} is based on a single accepting set; and 
2) the method using a reward function based on the accepting frontier function \cite{HAK2019,HKAKPL2019}. 
For the three methods, we use Q-learning\footnote{We employ Q-learning here but any algorithm that maximizes the discounted expected reward can be applied to our proposed method.} 
with an epsilon-greedy policy. 
The epsilon-greedy parameter is given by $ \frac{0.95}{n_t(s^{\otimes})}$, where $n_t(s^{\otimes})$ is the number of visits to state $s^{\otimes}$ within $t$ time steps \cite{Singh1998}, so that it will gradually reduce to 0 to learn an optimal policy asymptotically.
We set the positive reward $r_p = 2$ and the discount factor $\gamma = 0.95$. 
The learning rate $\alpha$ varies in accordance with {\it the Robbins-Monro condition}. 
We train the agent in 10000 iterations and 1000 episodes for 100 learning sessions.

\begin{figure}[tbp]
 \centering
 \begin{tabular}{c}
  \begin{minipage}{0.5\hsize}
     \centering
     \includegraphics[bb=0 0 461 346, height = 3.3cm, width=4.2cm]{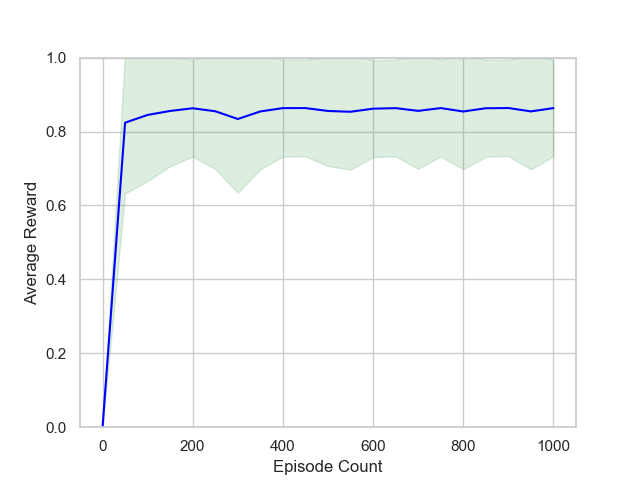}
 \end{minipage}

 \begin{minipage}{0.5\hsize}
   \centering
   \includegraphics[bb=0 0 461 346, height = 3.3cm, width=4.2cm]{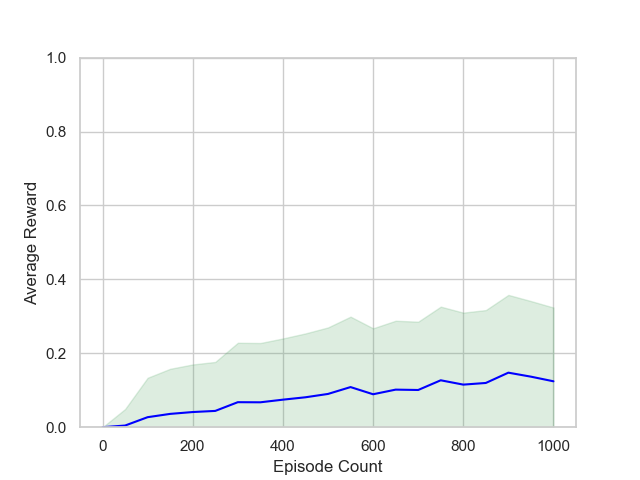}
 \end{minipage}
\end{tabular}
 \caption{The mean of average reward in each episode for 100 learning sessions obtained from our proposed method (left) and the method using tLDBA (right). They are plotted per 50 episodes and the green areas represent the range of standard deviations. }
 \label{result}
\end{figure}

\begin{figure}[tbp]
	\centering
	\begin{tabular}{c}
		\begin{minipage}{0.499\hsize}
		\centering
			\includegraphics[bb=0 0 341 256, height = 2.7cm,
			width=3.5cm]{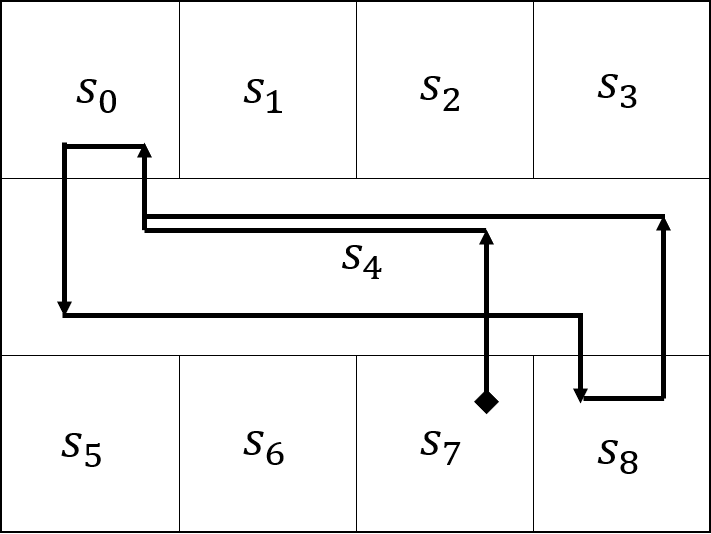}
		\end{minipage}
		\begin{minipage}{0.499\hsize}
			\centering
			\includegraphics[bb=0 0 341 257, height = 2.7cm,
			width=3.5cm]{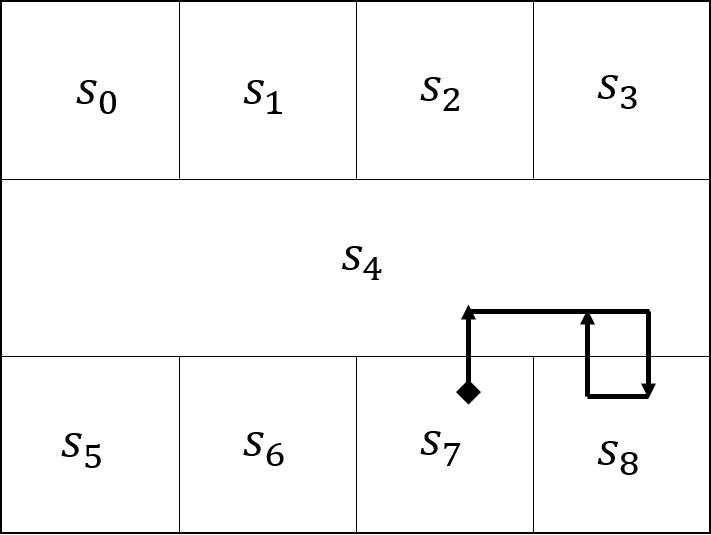}
		\end{minipage}
	\end{tabular}

	\caption{The optimal policy obtained from our proposed method (left) and the method in \cite{HAK2019, HKAKPL2019} (right).}
	\label{optimal}
\end{figure}

\subsection*{Results}
\textit{1) }
Fig.\ \ref{result} shows the average rewards obtained by our proposed method and the case using a tLDBA $B^{\prime}_{\varphi}$ converted from $\varphi$, respectively.
Both methods eventually acquire an optimal policy satisfying $\varphi$. As shown in Fig.\ \ref{result}, however, our proposed method converges faster. This is because the order of entrances to the rooms $s_0$ and $s_8$ is determined according to the tLDBA. 
Moreover, the number of transitions with a positive reward in $\bar{B}_\varphi$ is larger than that in $B_\varphi'$.

\textit{2) }
 We use the accepting frontier function \cite{HAK2019,HKAKPL2019} for the tLDGBA $Acc : \delta \times 2^{\delta} \rightarrow 2^{\delta} $. Initializing a set of transitions $ \mathbb{F} $ with the set of the all accepting transitions in $B_{\varphi}$, the function receives the transition $(x, \sigma, x^{\prime})$ that occurs and the set $\mathbb{F}$. If $(x, \sigma, x^{\prime})$ is in $\mathbb{F}$, then $Acc$ removes the accepting sets containing $(x, \sigma, x^{\prime})$ from $\mathbb{F}$. For the product MDP of the MDP $M$ and the tLDGBA $B_{\varphi}$, the reward function is based on the removed sets of $B_{\varphi}$.
 Then, we synthesize a positional policy on the product MDP derived from the tLDGBA $B_\varphi$.

\begin{figure}[tbp] 
	\centering
		\includegraphics[width=0.9\linewidth]{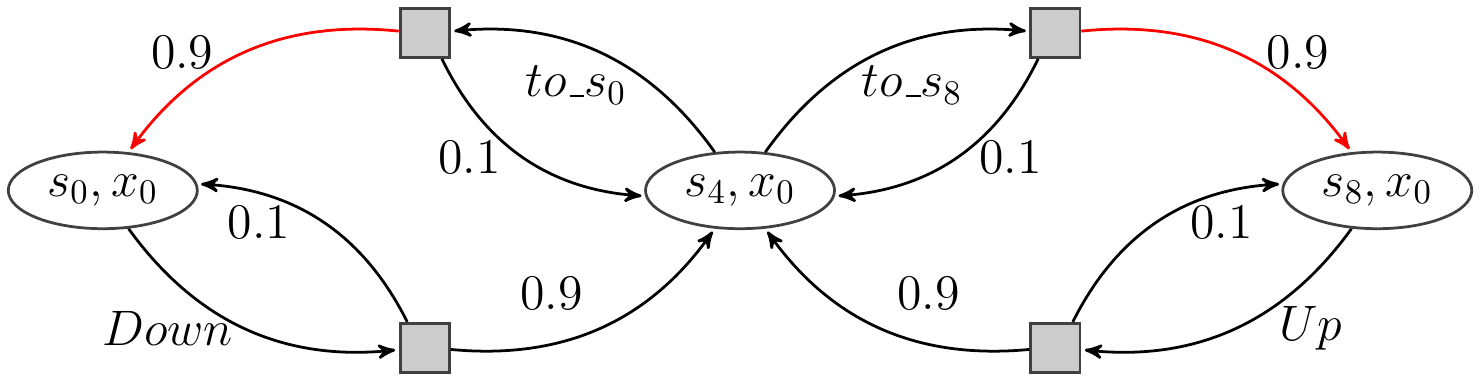}
	\caption{A part of the product MDP of the MDP shown in Fig.\ \ref{Grid1} and the tLDGBA shown in Fig.\ \ref{automaton}, where each transitions is labeled with either an action or the transition probability.}
	\label{part_product}
\end{figure}

Fig.\ \ref{optimal} shows the optimal policies obtained by our proposed method and the method in \cite{HAK2019,HKAKPL2019}, respectively.
The policy obtained by the method with the accepting frontier function fails to satisfy the LTL specification\footnote{We obtain the same result even with a state-based LDGBA.} 
because it is impossible that the transitions labeled with $\{ a \}$ and $\{ b \}$ occur from $s_4$ infinitely often by any positional policy on the product MDP with $B_{\varphi}$ shown in Fig.\ \ref{automaton}. 
More specifically, as shown in Fig.~\ref{part_product}, the agent cannot take each accepting transition  colored with red by any deterministic stationary action selection at the product state $(s_4,x_0)$. 
In our proposed method, 
the augmented tLDGBA includes the information of the (path-dependent) domain of the accepting frontier function explicitly as memory vectors, 
which enables us to synthesize a positional policy satisfying $\varphi$ on the product MDP.

\section{Conclusion}
The letter proposed a novel RL-based method for the synthesis of a control policy for an LTL specification using an augmented tLDGBA. 
The proposed method improved the learning performance compared to existing methods.
It is future work to investigate a method that maximizes the satisfaction probability.

\ifCLASSOPTIONcaptionsoff
  \newpage
\fi



\bibliographystyle{IEEEtran}
\bibliography{tex_resubmit/bibdata}
\end{document}